\documentclass[12pt]{iopart}
\usepackage{graphicx}
\usepackage{float}
\usepackage{bm}

\usepackage[utf8]{inputenc}
\usepackage[T1]{fontenc}
\usepackage{mathptmx}
\usepackage{etoolbox}
\usepackage{xcolor}
\usepackage{dcolumn}

\usepackage{bbold}
\usepackage{float}
\usepackage{booktabs}
\usepackage{multirow}
\usepackage{amssymb}
\usepackage{textcomp}
\usepackage{gensymb}
\usepackage{hyperref}
\usepackage{cite}

\begin{document}

\title[]{Mapping Charge-Transfer Excitations in Bacteriochlorophyll Dimers from First Principles}

\author{Zohreh Hashemi$^1$, Matthias Knodt$^1$, Mario R. G. Marques$^1$, Linn Leppert$^{1,2}$}
\address{$^{1)}$Institute of Physics, University of Bayreuth, Bayreuth 95440, Germany,

$^{2)}$MESA+ Institute for Nanotechnology, University of Twente, 7500 AE Enschede, The Netherlands}
\ead{l.leppert@utwente.nl}

\date{today}

\begin{abstract}
Photoinduced charge-transfer excitations are key to understand the primary processes of natural photosynthesis and for designing photovoltaic and photocatalytic devices. In this paper, we use Bacteriochlorophyll dimers extracted from the light harvesting apparatus and reaction center of a photosynthetic purple bacterium as model systems to study such excitations using first-principles numerical simulation methods. We distinguish four different regimes of intermolecular coupling, ranging from very weakly coupled to strongly coupled, and identify the factors that determine the energy and character of charge-transfer excitations in each case. We also construct an artificial dimer to systematically study the effects of intermolecular distance and orientation on charge-transfer excitations, as well as the impact of molecular vibrations on these excitations. Our results provide design rules for tailoring charge-transfer excitations in Bacteriochloropylls and related photoactive molecules, and highlight the importance of including charge-transfer excitations in accurate models of the excited-state structure and dynamics of Bacteriochlorophyll aggregates.
\end{abstract}

\newpage
\section{Introduction}
\label{intro}
Photoinduced charge-transfer excitations are of central importance to the primary processes of natural photosynthesis and for photovoltaic and photocatalytic applications \cite{wahadoszamen_role_2014, zoppi_charge-transfer_2018}. In organic semiconductors, charge-transfer excitations are believed to be important intermediates between excited states localized on donor molecules and charge-separated electron-hole states on acceptor and donor units, respectively, even though the exact mechanism of charge-separation is debated \cite{muntwiler_coulomb_2008, ohkita_charge_2008, pensack_barrierless_2009, lee_charge_2010, bakulin_role_2012, caruso_long-range_2012, murthy_mechanism_2012, yost_electrostatic_2013, jakowetz_what_2016, lee_higher-energy_2017}. In photosynthesis, the efficient conversion of solar energy into Chem. energy is achieved by structurally complex aggregates of Bacteriochlorophylls (BCL), Chlorophylls, and other pigment molecules embedded in transmembrane proteins that modulate their structure and function. These pigment-protein complexes form light-harvesting complexes and reaction centers that are responsible for photon absorption, excitation-energy transfer, and charge-separation. Their main operating principles are well-understood due to a wealth of crystallographic and spectroscopic studies complemented by numerical modelling using semi-empirical and first-principles approaches \cite{Jordanides2001,camara-artigas_interactions_2002,Jonas1996,Schlau-Cohen2012,Rancova2016, mirkovic2017light,Niedringhaus2018,kawashima_energetic_2018,Kavanagh2020,cupellini2020successes}.
\begin{figure}[htb]
\centering
\includegraphics[width=0.6\textwidth]{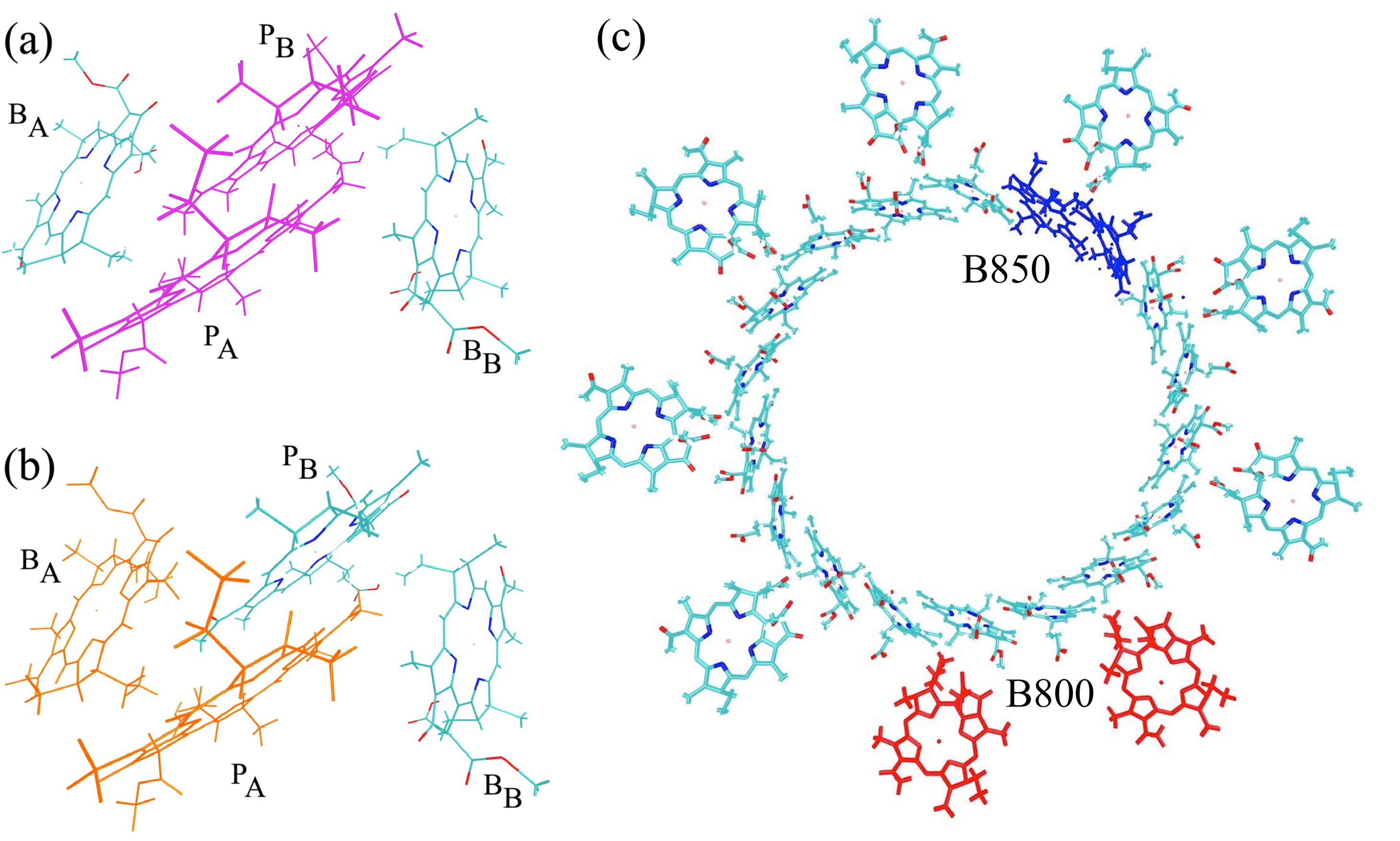}
\caption{Crystal structure of BCL aggregates in the reaction center (RC) and light-harvesting II (LHII) complex of the purple bacteria \textit{Rhodobacter sphaeroides} and \textit{Rhodoblastus acidophilus}, respectively. Dimers of BCLs are highlighted in color using (a) pink for the \textit{special pair} $P_A$ -- $P_B$, (b) orange for the A branch dimer P$_A$ -- B$_A$, (c) red for a dimer from the B800 and blue for a dimer from the B850 ring of the LHII complex. Hydrogen atoms are omitted for clarity.}
\label{fig1}
\end{figure}

In purple bacteria, charge separation occurs in reaction centers (RCs) comprising a hexameric aggregate of four BCLs and two Bacteriopheophytins, tightly surrounded by several protein chains \cite{wraight_absolute_1974, kirmaier_picosecond-photodichroism_1985, zinth_first_2005}. The primary four BCL molecules of this reaction center are shown in Figure~\ref{fig1}a, highlighting the so-called \textit{special pair} (SP), a strongly-coupled dimer of BCLs called $P_A$ -- $P_B$ in the following. Charge separation in the bacterial RC is initiated by a series of energy- and charge-transfer excitations that involve the SP and proceed along the \textit{A branch}, the photoactive of the two pseudo-symmetric branches the RC consists of \cite{ma_vibronic_2018,Niedringhaus2018,Ma2019,policht_hidden_2022}. In Figure~\ref{fig1}b, we have highlighted the A-branch dimer $P_A$ -- $B_A$ that has been speculated to be involved in the primary charge-separation step, although this assignment is debated in the literature \cite{van_brederode_efficiency_1998, zhou_probing_1998, lin_excitation_1998, huang_cofactor-specific_2012}. Excitation energy reaches the RC through a cascade of excitation-energy transfer processes that are initiated in the light harvesting II (LHII) complex, consisting of two rings of BCL molecules dubbed B850 and B800, respectively, and shown in Figure~\ref{fig1}c. Neighboring BCLs in the B800 ring are only weakly coupled and excitation-energy transfer is well-described by Förster dipole-dipole coupling \cite{fassioli_photosynthetic_2014}. In the B850 ring, neighboring BCL molecules are closer and intermediate between the weakly coupled B800 and the strongly coupled special pair BCLs.

The excited states that are believed to be responsible for excitation energy transfer in and between the light-harvesting complexes and the RC, are commonly thought of as Frenkel-like excitons that are spatially relatively localized on one or two BCL molecules \cite{Jang2018}. Semi-empirical models based on Frenkel-excitons Hamiltonians have played an important role in modelling the excitation-energy and charge-transfer dynamics in large photosynthetic pigment-protein complexes \cite{VanderVegte2015, Curutchet2016,thyrhaug_identification_2018}. However, for a reliable and predictive representation of the electronic coupling between adjacent pigments, charge-transfer excitations need to be included in these model Hamiltonians \cite{voityuk_fragment_2014, voityuk_interaction_2015, Curutchet2016, Li2017a, sen_understanding_2021}, calling for accurate first-principles calculations of such excitations.

For computationally efficient first-principles methods such as time-dependent density functional theory (TDDFT), charge-transfer excitations were long considered a major challenge due to their inherently nonlocal nature, i.e., the spatial separation of the occupied and virtual orbitals contributing to these excitations \cite{Dreuw2004}. TDDFT with optimally-tuned range-separated hybrid functionals is a viable solution to this problem, and has been used to predict excited states of molecular systems and solids with great success \cite{Refaely-Abramson2011, Refaely-Abramson2012, Korzdorfer2012, Refaely-Abramson2013, DeQueiroz2014, manna_quantitative_2018, Wing2020a}. In these exchange-correlation functionals, the presence of long-range exact exchange leads to asymptotically correct potentials. Additionally, a parameter controlling the range-separation of exact and semilocal exchange can be used to tune the energies of the highest occupied and the lowest unoccupied orbitals to correspond to the negative of the ionization potentials and the electron affinity, respectively, within the conceptual framework of generalized Kohn-Sham \cite{Seidl1996}. Both conditions are crucial for accurately capturing charge-transfer excitations within linear-response TDDFT \cite{Kuemmel2017} and have been extended to solvated molecular systems \cite{DeQueiroz2014, bhandari_fundamental_2018} and extended solids \cite{Wing2020a}. 

An alternative approach for calculating charge-transfer excitations of molecules and solids is the $GW$+Bethe-Salpeter Equation ($GW$+BSE) approach \cite{Rohlfing1998, Rohlfing2000}. While this method was initially primarily applied to solids, recent years have witnessed a multitude of studies that have demonstrated the accuracy and predictive power of the $GW$+BSE method for small molecules \cite{VanSetten2015, Bruneval2015, Rangel2017a} and larger molecular complexes \cite{Duchemin2012, duchemin_resonant_2013, Sharifzadeh2013, Blase2018, forster_quasiparticle_2022}. In particular, we \cite{Hashemi2021a} and others \cite{forster_quasiparticle_2022} benchmarked the accuracy of the $GW$+BSE approach against experiment and wavefunction-based methods and found excellent agreement for the $Q_y$ and $Q_x$ excitations of a range of BCL and Chlorophyll molecules. We showed that both eigenvalue self-consistent $GW$ calculations and \textit{one-shot} $G_0W_0$ calculations where the zeroth-order single-particle Green's function $G_0$ and screened Coulomb interaction $W_0$ were constructed from a DFT eigensystem obtained with an optimally-tuned range-separated hybrid functional lead to the best results. TDDFT with an optimally-tuned hybrid-functional performed slightly worse and tended to overestimate the energy of the $Q_y$ excitations, in agreement with previous studies \cite{Duchemin2012}. 

In this article, we report a systematic first-principles study of charge-transfer excitations in BCL dimers - the smallest structural units in which excitations with intermolecular charge-transfer character can be observed. Two types of BCL dimers constitute our model systems in this study: The first class of dimers (discussed in Section~\ref{sec:real-dimer}) is extracted from the LHII complex and RC of purple bacteria. We note that the excitation energies that we calculate for these systems will be different from those \textit{in vivo}, where electrostatic and dielectric effects of the protein environment and coupling with other pigments leads to different excitations and affects locally excited and charge-transfer states differently \cite{Jang2018,volpert_delocalized_2022,aksu2019explaining,sirohiwal2020protein}. Our goal is to elucidate the factors that determine the energy and character of these excitations, in particular their mixing with the coupled $Q_y$ and $Q_x$ excitations of the dimers. We treat these dimers as representative of four different regimes of intermolecular coupling resulting in distinct charge-transfer properties: 1. The B800 dimer is very weakly coupled with $Q_y$ and $Q_x$ excitations resembling those of the monomeric units and high-energy charge-transfer excitations due to vanishing orbital overlap. 2. The A-branch dimer is more strongly coupled and exhibits one charge-transfer excitation corresponding to electron transfer from $P_A$ to $B_A$. We use the notation $P_A^+B_A^-$ to indicate the direction of charge-transfer in the following. This charge-transfer excitation is $\sim$0.4\,eV higher in energy than the coupled $Q_x$ excitations. 3. The B850 dimer is even more strongly coupled. The lowest-energy charge-transfer excitation mixes with the coupled $Q_x$ excitations and another charge-transfer state appears at slightly higher energies. 4. Finally, the special pair SP is the most strongly coupled case with three charge-transfer excitations mixing with the coupled $Q_x$ excitations. Additionally, we construct an artificial BCL dimer and systematically study the effects of intermolecular distance and orientation on charge-transfer excitations. We also estimate the effect of molecular vibrations on charge-transfer excitations. We do this by calculating the vibrational normal modes of a dimeric system and determining the change of excitation energies for structures distorted along normal modes. This allows us to identify vibrational modes with pronounced effects on charge-transfer excitations. Finally, we comment on differences and similarities between TDDFT with an optimally-tuned range separated hybrid functional and the $GW$+BSE approach.

\section{Computational Methods}
\label{sec:methods}
\subsection{First-Principles Methods and Computational Details}\label{sec:computational}
For all calculations reported in this article, we used TDDFT as implemented in \textsc{turbomole} version 7.5 \cite{ahlrichs1989electronic} and the $GW$+BSE approach as implemented in \textsc{molgw} version 3.0 \cite{bruneval2016molgw}. Briefly, in the linear-response formulation of both methods the excitation energies $\Omega_n$ can be obtained by solving the matrix eigenvalue equation $C\mathbf Z = \Omega_n^2 \mathbf Z$, where $C$ is
\begin{equation}
\label{eq:Casida}
C_{ij\sigma, kl\tau} = (\varepsilon_{i\sigma} - \varepsilon_{j \sigma})^2 \delta_{ij}\delta_{jl}\delta_{\sigma \tau} + 2 \sqrt{\varepsilon_{i\sigma} - \varepsilon_{j\sigma}} \sqrt{\varepsilon_{k\tau} - \varepsilon_{l\tau}} K_{ij \sigma, kl \tau}    
\end{equation}
and the indices $i, k$ refer to occupied, $j, l$ to virtual orbitals and $\sigma, \tau$ to spin-indices. Differences between TDDFT and the $GW$+BSE approach enter Equation~\ref{eq:Casida} in two distinct ways: 1. Through the differences between virtual and occupied orbital energies $\varepsilon_{i\sigma} - \varepsilon_{j \sigma}$ which are obtained from a (generalized) Kohn-Sham calculation in TDDFT and from the $GW$ approach in $GW$+BSE. 2. Through the kernel matrix element $K_{ij \sigma, kl \tau}$, which depends on the exchange-correlation kernel $f_{xc,\sigma}$ - the functional derivative of the exchange-correlation potential - in TDDFT, and on the screened Coulomb interaction $W$, typically evaluated in the random phase approximation and at zero frequency, in the BSE approach \cite{Onida2002, blase2018bethe,blase2020bethe,marques2004time}. 

Here we use the optimally-tuned range-separated hybrid functional $\omega$PBE for our TDDFT calculations. We use a range-separation parameter $\omega$=0.171\,$a_0^{-1}$, based on tuning for a single BCL~$a$ molecule performed by Schelter \textit{et al.} \cite{schelter2019assessing}. The optimal-tuning procedure follows the recipe by Stein \textit{et al.} and ensures that the HOMO eigenvalue corresponds to the ionization potential and the LUMO eigenvalue corresponds to the electron affinity of the molecule \cite{Stein2010}. We do not perform a new tuning procedure for the dimers for general reasons: Using the same $\omega$ for each dimer allows us to compare the electronic and excited state structure of these systems on the same footing. Furthermore, optimal tuning of conjugated systems of increasing size leads to artificially low values of $\omega$ and, thus, a dominance of semilocal exchange at long range, which deteriorates the description of charge-transfer excitations \cite{Korzdorfer2011a, DeQueiroz2014}. 

For our $GW$+BSE calculations we use a "one-shot" $G_0W_0$ approach in which we construct the zeroth-order single-particle Green's function $G_0$ and the screened Coulomb interaction $W_0$ from DFT eigenvalues and eigenfunctions calculated using the same $\omega$PBE as described above. This approach leads to excellent agreement with experimental excitation energies and reference values from wavefunction-based methods for a range of BCL and Chlorophyll molecules \cite{Hashemi2021a}. Range-separated hybrid functionals have been shown to lead to accurate charge-transfer excitations for larger molecular complexes as well \cite{Duchemin2012, baumeier_frenkel_2012}. In all calculations we used a def2-TZVP basis set, and the frozen core and resolution-of-the-identity approximations (with the DeMon auxiliary basis set \cite{Godbout1992}). We did not apply the Tamm-Dancoff approximation in any of the results reported in this paper. In our $G_0W_0$ calculations, we used the optimized virtual subspace method by Bruneval with an aug-cc-pVDZ basis set for the reduced virtual orbital subspace \cite{Bruneval2016a}. With these settings, our excitation energies are converged to within 40\,meV. Further details on our convergence tests can be found in Section~\ref{sec:convergence} and in the Supplemental Material (SM). 

\begin{figure}[htb]
\centering
\includegraphics[width=7cm]{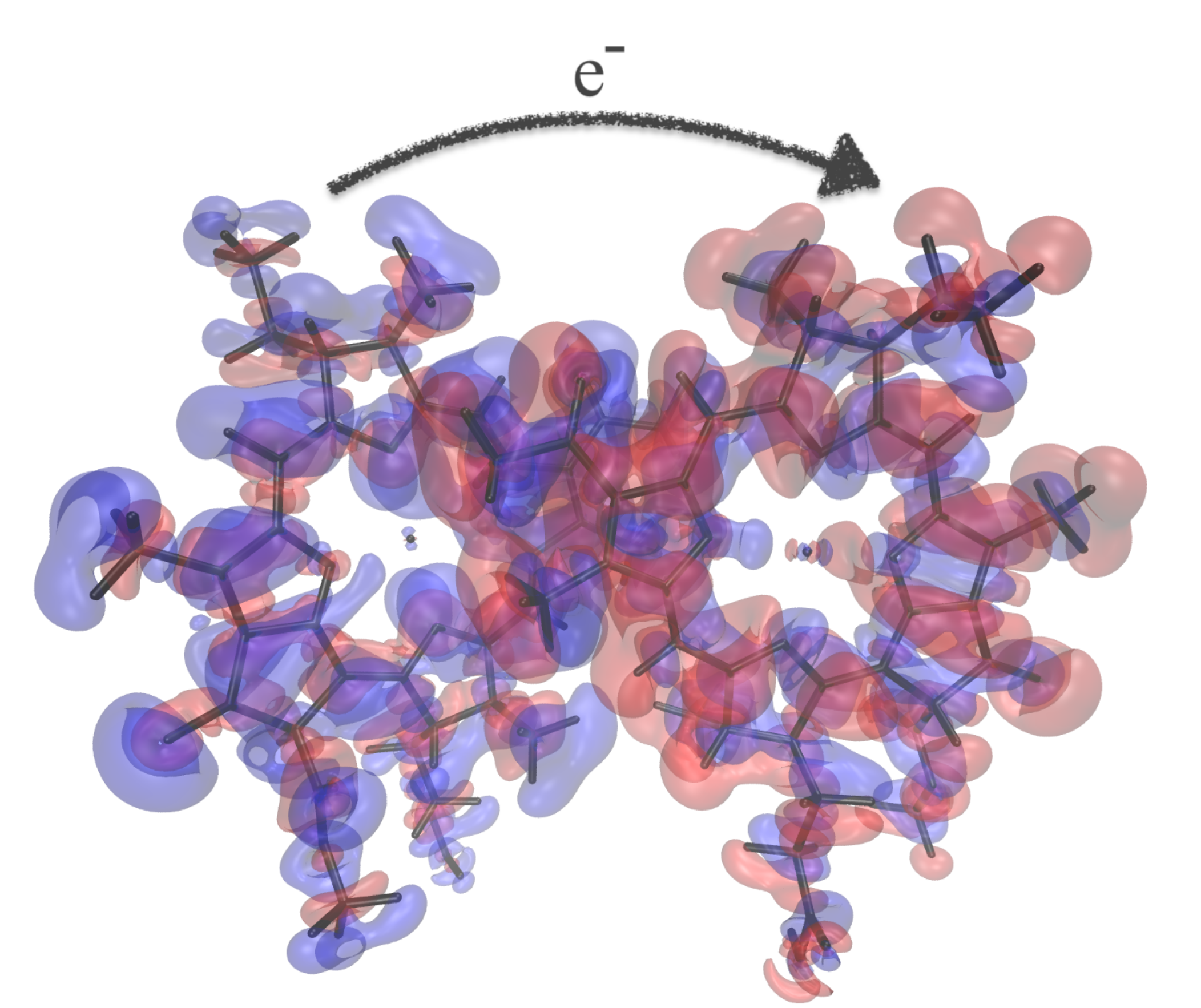}
\caption{Schematic isosurface picture of the difference density of an excited state. The red and blue areas correspond to regions of positive and negative density, respectively. We integrate the difference density over volumes associated with each molecule to quantify how much charge is transferred as a result of the excitation (see main text).}
\label{dd}
\end{figure}

For evaluating the character of the excited states, we calculated their transition and difference densities which are both derived from the density matrix $\gamma^{ii}(r,r')=N\int\Psi^{i}(r,r_2,r_3,...,r_n)\Psi^{i}(r',r_2,r_3,...,r_n)dr_2...dr_n$, where $N$ is the number of electrons and $\Psi^{i}$ is the generalized Kohn-Sham excited-state wavefunction, here constructed from a sum of Slater determinants of generalized Kohn-Sham orbitals with coefficients from linear-response TDDFT or the BSE. The ground state density is $n^0(r)= \gamma^{00} (r,r')$. The density of excited state $i$ is $n^i(r)= \gamma^{ii} (r,r')$. The difference density is obtained by subtracting $n^i$ from $n^0$, and allows to visualize the change of density upon excitation of the system into excited state $i$, as schematically shown in Figure~\ref{dd}. The transition density is obtained as the diagonal part of the density matrix for a transition from the ground state into an excited state $i$ $\rho^{0i}(r)= \gamma^{0i} (r,r')$, and is particularly useful for determining the interaction strength of electronic transitions with light and efficiencies of excitation energy transfer. For charge-transfer excitations there is no overlap between ground and excited state, thus the transition density vanishes. Therefore, the difference density is a more useful tool for visualizing charge-transfer excitations and quantifying their charge-transfer character \cite{Plasser2014}. Note that for the identification of charge-resonance excitations, i.e., excitations without a net charge transfer which can be described as linear combinations of forward and backward charge transfers of equal weight, direct analysis of the transition density matrix is a more useful tool \cite{Plasser2012}. In the following, we define charge-transfer excitations as those with a net shift of charge with respect to the ground state.

To quantify the magnitude of charge transfer we integrated over subsystem difference densities. For this purpose, we subdivided the volume containing the difference densities of the dimer into subsystem volumes, each containing one pigment. Our aim is to assign each grid point of the difference-density grid to its closest pigment molecule. For achieving this, we used the distances between grid points and each molecule's atomic coordinates (including hydrogen atoms), as previously done in Ref.~\cite{volpert_delocalized_2022}.

Finally, to obtain a mode-resolved picture of the effect of thermally-activated vibrations (Section~\ref{sec:vib}), we relaxed a dimer structure using the B3LYP approximation for the exchange-correlation functional and def2-TZVP basis set, and evaluated its normal modes and frequencies. Using the harmonic approximation, we related the amplitude of these normal modes with the thermal energy of a molecule. Thus, we distorted the dimer structure along its lowest-frequency normal modes at a temperature of 300\,K. In this manner, we generated 60 distortions of the dimer, that we then studied using TDDFT calculations using the $\omega$PBE functional. All these calculations were performed using the tools provided in the \textsc{turbomole} package.

\subsection{Convergence of $G_0W_0$+BSE calculations}\label{sec:convergence}
We carefully tested that our $GW$+BSE results are converged. Due to the large size of a BCL dimer, featuring more than 300 electrons, the calculation of the $GW$ self-energy which requires summation over virtual states is computationally demanding. We therefore used the optimized virtual subspace method implemented in the \textsc{molgw} code, in which a reduced virtual orbital subspace represented by a comparably small basis set is used to evaluate the $GW$ self-energy \cite{Bruneval2016a}.
\begin{figure}[htb]
\centering
\includegraphics[width=14cm]{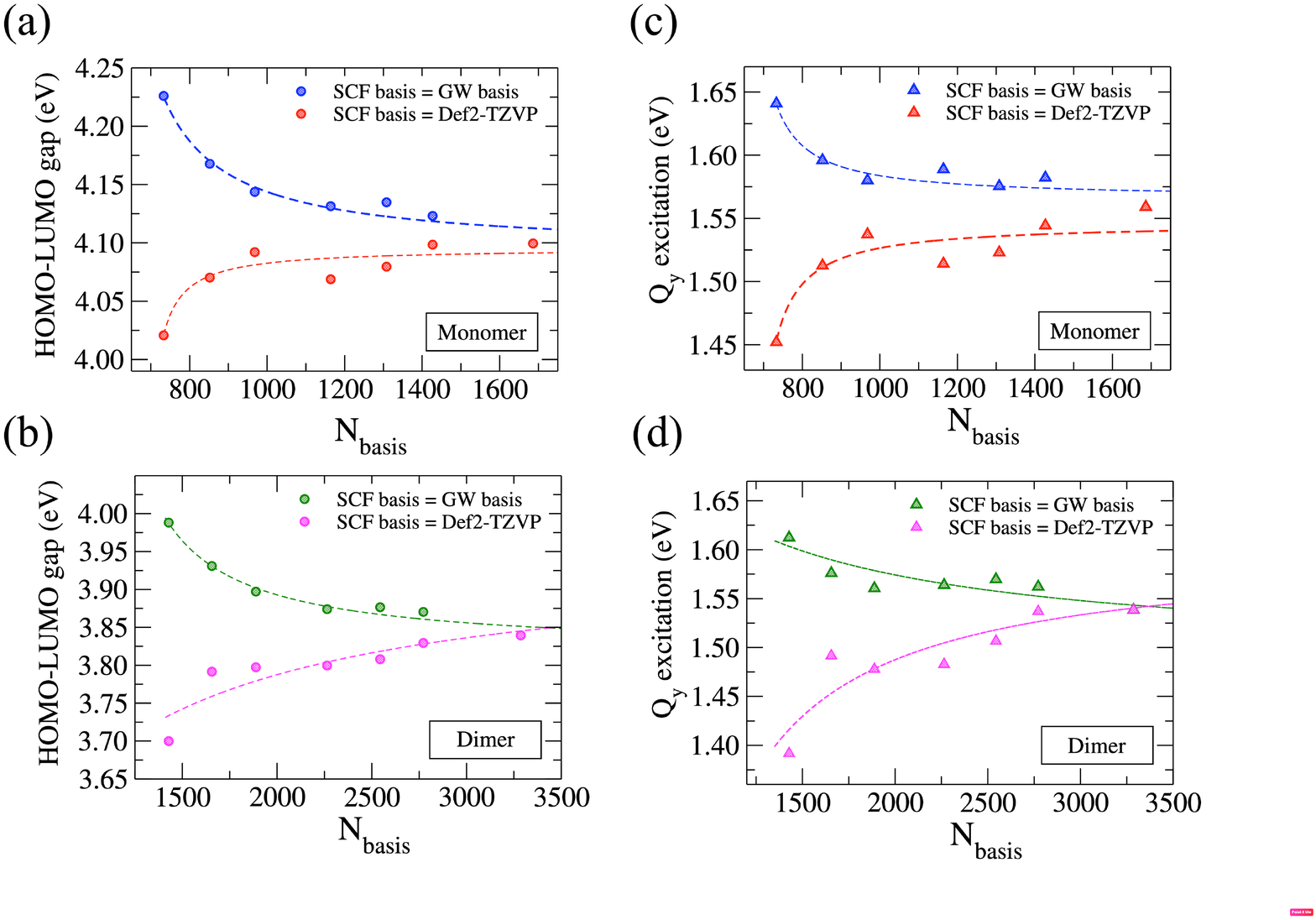}
\caption{Convergence of $GW$ (a) HOMO-LUMO gap and (c) energy of the first excited state of BCL$~a$ monomer as a function of the number of basis functions. Blue data points correspond to calculations in which the same basis set is used for the occupied orbitals and the virtual subspace. Red points correspond to calculations using the optimized virtual subspace method. Lines are fits to these data points. Convergence of the HOMO-LUMO gap and energy of the first excited state is shown in panel (b) and (d) for the B850 dimer, respectively. Here, green corresponds to using the same basis set for the occupied orbitals and the virtual subspace and pink to calculations using the optimized virtual subspace method.}
\label{fig3}
\end{figure}

We start by testing the convergence of the HOMO-LUMO gap, and the $Q_y$ and $Q_x$ excitations of a BCL~$a$ monomer with respect to basis set size without the optimized virtual subspace method (Table S1). In agreement with our previous results \cite{Hashemi2021a}, we find that the def2-TZVP basis set deviates by less than 10\,meV from the considerably larger aug-cc-pVTZ basis. We proceeded by calculating the convergence of the $Q_y$ and $Q_x$ excitations of the BCL monomer as a function of the number of virtual orbitals $N_{virt}$ included in the evaluation of the $GW$ self-energy using the def2-TZVP basis (Figure S1). We find that for $N_{virt} = 500$ both excitations are converged to within 80\,meV from the limit of infinite $N_{virt}$. Based on these findings we continued by evaluating the effect of using a smaller basis set for the virtual subspace \cite{Bruneval2016a}. The results for the HOMO-LUMO gap and the $Q_y$ excitation are plotted in Figure~\ref{fig3}a and c, and show that the optimized virtual subspace method leads to an underestimation of the HOMO-LUMO gap and the $Q_y$ excitation energy as compared to the \textit{conventional} method in which the same basis set is used for all orbitals. We find that using the aug-ccpVDZ basis for the optimized virtual subspace in conjunction with $N_{virt} = 500$ leads to a fortuitous error cancellation and results in a HOMO-LUMO gap and $Q_y$ and $Q_x$ excitation energies that are within less than 50\,meV of the results obtained with the conventional method and $N_{virt} \rightarrow \infty$ (Figure S2).

For the dimer, we therefore chose $N_{virt} = 1000$ and the same strategy for determining the optimized virtual subspace. We find very similar results for the convergence of the HOMO-LUMO gap and the first bright coupled $Q_y$ excitation shown in Figure~\ref{fig3}b and d. All $GW$+BSE results reported in this paper are therefore based on calculations using the def2-TZVP basis set for the occupied orbitals and the aug-ccpVDZ basis for the optimized virtual subspace.

\subsection{Construction of the Model Systems}
We constructed our model systems from the x-ray crystallographic structures of the purple bacteria \textit{Rhodobacter sphaeroides} (structure ID $1M3X$ in the Protein Data Base) \cite{camara2002interactions},  \textit{Rhodoblastus acidophilus} (structure ID $1NKZ$ \cite{papiz2003structure}). In all structures, we replaced the phytyl tail with hydrogen. Hydrogen atoms not resolved in the experimental crystal structure were added using \textsc{avogadro} and their positions were optimized while keeping the rest of the structure fixed. These geometry optimizations were performed using \textsc{turbomole} and the B3LYP exchange-correlation functional. The reaction center dimers $P_A$ -- $P_B$ and $P_A$ -- $B_A$ (Figures~\ref{fig1}a and b) were constructed using structure $1M3X$ while the B800 and B850 ring dimers (Figure~\ref{fig1}c) were extracted from $1NKZ$. These molecules correspond to ID numbers $BCL307$ and $BCL309$ for the B800, and $BCL302$ and $BCL303$ for the B850 ring. Note that the resolution of X-ray crystal structures of protein complexes is often not high enough to resolve the internal structure of the chromophores. Therefore, internal coordinates (such as bond lengths) can be unreliable, and care should be taken when comparing calculations to experiment \cite{Dreuw2010}. We illustrate this effect in Figure~S3, where we compare TDDFT spectra based on the X-ray crystal structures with those based on constrained relaxations of the B800 and the B850 dimers. Geometry optimization leads to significant changes in the excitation energies, in particular a blueshift of the Q$_y$ excitation and charge-transfer states. Nonetheless, in Section 3.1, we present results based on unrelaxed crystal structures, since qualitatively the results are the same.

To study the effect of structure in more detail, we additionally constructed an artificial dimer consisting of two exactly equivalent relaxed BCL~$a$ molecules (using molecule $P_A$) that we initially oriented in the same way as the special pair dimer $P_A$ -- $P_B$ by aligning their transition dipole moments (as calculated with TDDFT) with those of $P_A$ and $P_B$, respectively. We are providing all relevant structure files necessary to reproduce the results of this article in the SM.

\section{Discussion and Results}
\subsection{Charge-Transfer Excitations in RC and LHII Dimers}
\label{sec:real-dimer}
\begin{figure}[htb]
\centering
\includegraphics[width=12cm]{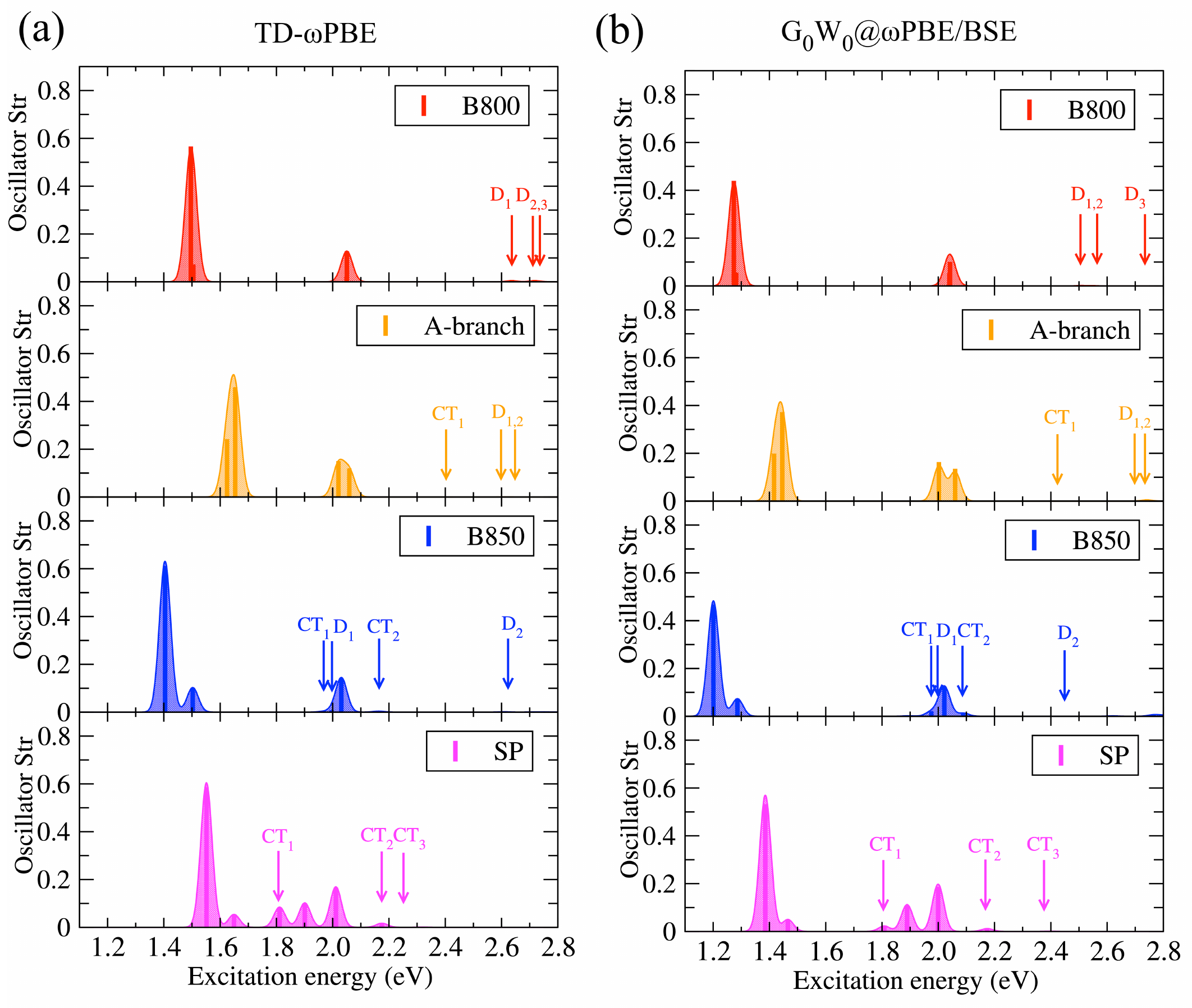}
\caption{Excitation spectrum of B800, A-branch, B850, and SP dimers using (a) TDDFT with $\omega$PBE and (b) the $G_0W_0$@$\omega$PBE+BSE approach. Arrows mark dark excitations without (D) and with (CT) charge-transfer character. The shaded areas are calculated by folding the excitation energies with Gaussian functions with a width of 0.08\,eV as a guide to the eye.}
\label{fig4}
\end{figure}

We start by comparing the excitation spectrum of the four dimeric systems shown in Figure~\ref{fig1}a-c using TDDFT and $GW$+BSE. The energies and oscillator strengths of the first 15 excitations of each system can be found in Table S3 and S4. The spectra are shown in Figure~\ref{fig4}a and b, respectively, and allow for several observations. First, we find that TDDFT and $GW$+BSE predict qualitatively very similar spectra. The most striking difference appears for the B800 dimer, for which the coupled $Q_y$ excitations calculated with TDDFT are $\sim$0.3\,eV higher in energy than with $GW$+BSE while all other excitations are at similar energies. This observation is consistent with our results for single BCL~$a$ molecules for which TDDFT with optimally-tuned $\omega$PBE consistently overestimates the $Q_y$ excitation energy by $\sim$0.3\,eV \cite{Hashemi2021a} and therefore leads to an underestimation of the $Q_y$ -- $Q_x$ energy difference as compared to experiment. Interestingly, this overestimation as compared to $GW$+BSE, while still present, is less pronounced for the other three dimers and seems to decrease with increasing intermolecular coupling. Our results are in qualitative agreement with previous calculations (see Table S5), but a comparison is complicated by the use of different structural models, exchange-correlation functionals and basis sets.

Second, we find several dark excitations for all four systems, predicted at very similar energies with TDDFT and the $GW$+BSE approach. We analyze the charge-transfer character of these excitations by calculating their difference densities and integrating over subsystem difference densities as described in Section~\ref{sec:computational}. The energy and character of these dark excitations considerably differs for our four dimers. For the B800 dimer, we find three dark excitations, E$_5$, E$_6$, and E$_7$, $\sim$0.7\,eV above the coupled Q$_x$ excitations which are almost degenerate. The difference densities (Figure~S4 and Table~\ref{table2}) do not indicate any charge-transfer character for these excitations - their charge distribution is primarily localized on only one BCL in each excitation, and looks similar to those of the monomeric system. Charge-transfer excitations can be found at around 3.0\,eV, consistent with the large distance of 20\,\AA{} between the B800 molecules, measured as the distance between their centers of masses.
\begin{table}[htb]
\centering
\begin{tabular}[t]{l|c|cccccc}
\toprule
   dimer  &  molecule label  & \multicolumn{5}{c}{charge distribution}\\
 &&E$_3$& E$_4$& E$_5$& E$_6$& E$_7$ \\
\midrule                
   \multirow{2}{*}{B800}& B307 & 0 &0 &0 &0 &0\\
                        & B309 & 0&0 &0 &0 &0 \\\midrule
   \multirow{2}{*}{A-branch}& P$_A$& 0&0 &-0.97 &0 &0\\
                        & B$_A$ & 0&0&0.97&0&0\\\midrule                       
   \multirow{2}{*}{B850}& B302 & -0.83& 0 & 0 & 0.86 &0\\
                        & B303 & 0.83& 0 & 0 & -0.86 &0\\\midrule
    \multirow{2}{*}{SP} & P$_A$ & -0.69&0 &0 &0.83 &-0.76\\
                        & P$_B$ & 0.69&0 &0 &-0.83 &0.76\\\midrule
\bottomrule
	\end{tabular}
	\caption{Difference density integrated over subsystem volumes. The first two excitations, i.e., E$_1$ and E$_2$, are not included since their difference densities integrate to zero in all studied systems.}
\label{table2}
\end{table}

The molecules $P_A$ and $B_A$ of the A-branch dimer are $\sim$13\,\AA{} apart, leading to stronger intermolecular coupling and the appearance of a charge-transfer state in the energy range considered here. Figure~\ref{fig4} shows that for this system the coupled Q$_y$ and Q$_x$ excitations are split and the first dark excitation is $\sim$0.3\,eV higher in energy than the second coupled $Q_x$ excitation. Contrary to the B800 dimer, this dark excitation has clear charge-transfer character (Table~\ref{table2}) and corresponds to $P_A^+B_A^-$. The character of the two following dark states is unchanged as compared to B800 apart from a redshift. 

In the B850 dimer with $\sim$ 11\AA$~$distance, the stronger intermolecular coupling leads to a further redshift of the dark excitations. We find that a dark state mixes with the coupled $Q_x$ excitations leading to charge-transfer character in E$_3$. The second charge-transfer excitation, E$_6$, appears $\sim$0.2\,eV above the first one, in the vicinity but energetically well-separated from the coupled $Q_x$ excitations.

The excitation spectrum of the special pair dimer SP is yet different. Due to the strong intermolecular coupling of the two molecules which are only 9\,\AA{} apart, three charge-transfer excitations appear at relatively low energies. The first one is lower in energy than the first coupled $Q_x$ excitation and corresponds to $P_A^+P_B^-$, whereas the other two are above the coupled $Q_x$ excitations and correspond to $P_A^-P_B^+$ and $P_A^+P_B^-$, respectively. Note that due to the overestimation of the coupled $Q_y$ excitations by TDDFT,  $GW$+BSE predicts the energy gap between the coupled Q$_y$ excitations and CT$_1$ to be twice as large as TDDFT. Nonetheless, since the qualitative features of all four excitation spectra and the charge-transfer character of all excitations is similar, we use TDDFT for all further calculations and report $GW$+BSE results in the SM. 

\subsection{Charge-Transfer Excitations in Artificial Dimer}
\label{sec:art-dimer}
\begin{figure} [htb]
\centering
\includegraphics[width=10cm]{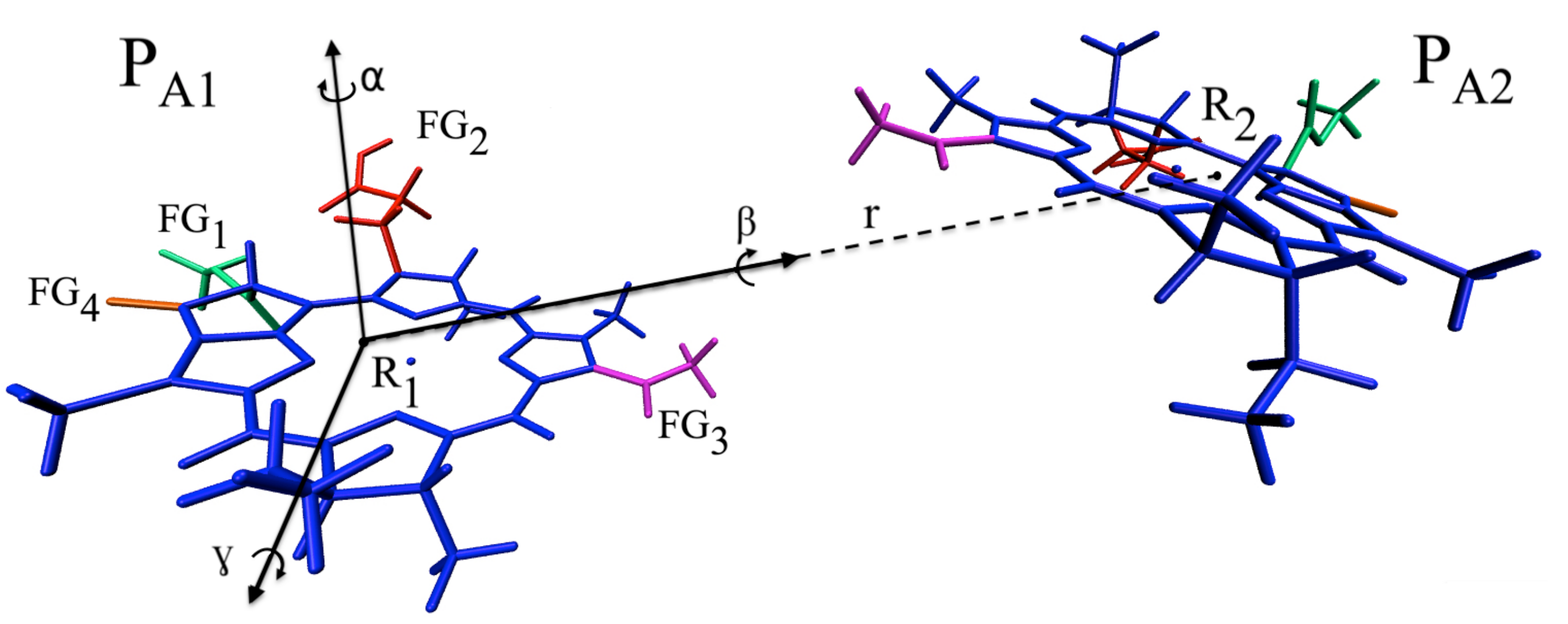}
\caption{Structure of artificial dimer based on two identical P$_A$ molecules. We highlight four functional groups FG$_1$ (in green), FG$_2$ (in red), FG$_3$ (in pink), and FG$_4$ (in orange). Hydrogen atoms are omitted for clarity.}
\label{fig5}
\end{figure}

The dimeric systems extracted from the RC and LHII crystal structures discussed in Section~\ref{sec:real-dimer}, differ in their distance, relative orientation, and the structural details of the two molecular subunits comprising the dimer. To disentangle these effects, we therefore proceeded by performing TDDFT calculations for an artificial dimeric system constructed as discussed in Section~\ref{sec:methods}. The structural parameters that define the distance and relative orientations of this dimer are shown in Figure~\ref{fig5}. We measure the distance between the molecules $r$ as the distance between their centers of masses $\mathbf R_1$ and $\mathbf R_2$, i.e., $r = |\mathbf r| = |\mathbf R_1 - \mathbf R_2$|. Their relative orientation is defined by three angles $\alpha$, $\beta$, and $\gamma$. The first angle, $\alpha$, is a rotation around the normal vector of the plane spanned by the $Q_y$ and $Q_x$ transition dipole moments of a single molecule, i.e., it is approximately perpendicular to the porphyrin-ring plane. The second rotation axis, associated with $\beta$, corresponds to $\mathbf r = \mathbf R_1 - \mathbf R_2$. The third rotation, $\gamma$, is around the axis given by the cross product of $\mathbf r$ and the normal vector of the $Q_y$ -- $Q_x$ plane. For our further discussion, we also distinguish between the four functional groups FG$_1$, FG$_2$, FG$_3$, and FG$_4$, highlighted in Figure~\ref{fig5}.

\begin{figure}[htb]
\centering
\includegraphics[width=\textwidth]{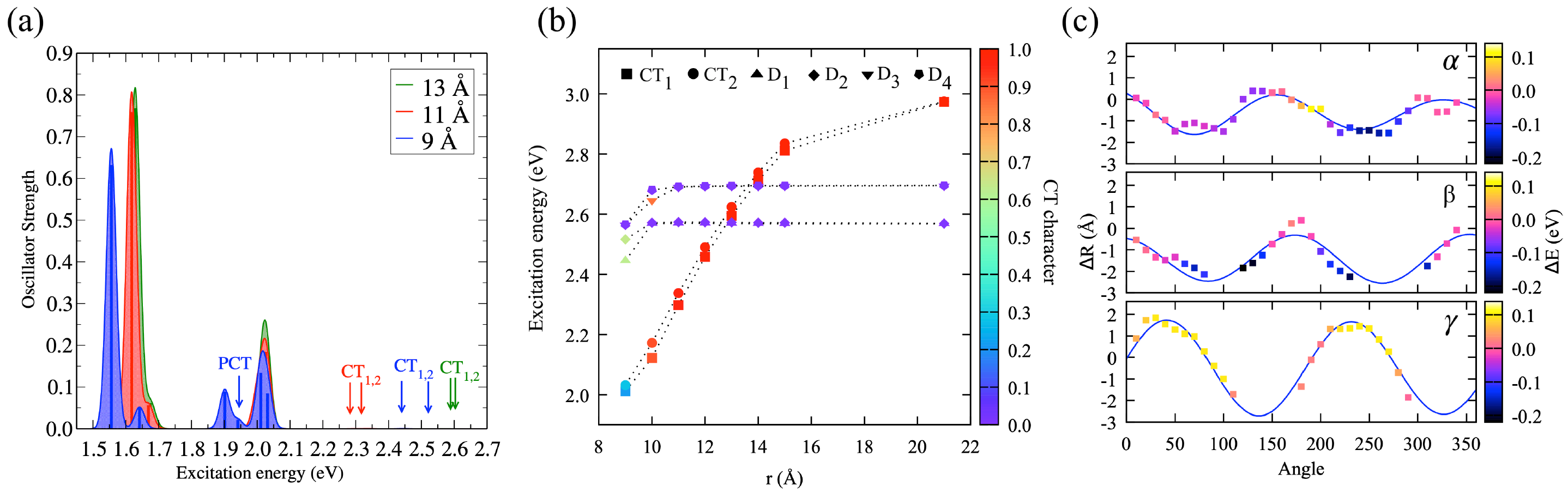}
\caption{(a) Absorption spectra of artificial dimer with $r = 9$\,\AA{} (blue), $r =11$\,\AA{} (red), and $r = 13$\,\AA{} (green). Arrows mark excitations with charge-transfer character. The shaded areas are calculated by folding the excitation energies with Gaussian functions with a width of 0.08\,eV as a guide to the eye. (b) The excitation energy of the first two charge-transfer (CT$_1$ and CT$_2$) excitations and the first four dark states (D$_1$-D$_4$) as a function of $r$. The color scale represents the charge-transfer character of each excitation based on the absolute value of the integrated subsystem difference densities. (c) $\Delta$R (see main text) as a function of the rotation angle $\alpha$ (top), $\beta$ (middle), and $\gamma$ (bottom). Blue lines are periodic fits and serve as a guide to the eye. The color scale corresponds to the change in energy $\Delta E$ of CT$_1$ as compared to the unrotated reference structure.}
\label{fig6}
\end{figure}

We start by investigating the effect of changing the distance $r$ between the molecules P$_{A1}$ and P$_{A2}$, fixing the relative orientation of the molecules such that it corresponds to the one found in the special pair dimer SP. Figure~\ref{fig6}a shows the excitation spectra of dimers separated by 9, 11, and 13\,\AA{}, corresponding to the center-of-mass difference found in the special pair SP, the B850 dimer, and the A-branch dimer of Section~\ref{sec:real-dimer}, respectively. Note that distances smaller than 9\,\AA{} are not possible for the artificial dimer due to overlap between the FG$_3$ functional groups. Decreasing the center-of-mass difference leads to a redshift and splitting of the coupled $Q_y$ excitations accompanied by a redistribution of oscillator strength between the two excitations, in accordance with expectations from Kasha's exciton theory \cite{kasha1965exciton}. The effect on the coupled $Q_x$ excitations cannot be discussed without also considering the higher-energy charge-transfer excitations. The latter are redshifted when going from 13\,\AA{} to 11\,\AA{}, and mix with the coupled $Q_x$ excitations at 9\,\AA{}, similar to the situation in the special pair dimer SP. The corresponding charge distributions based on subsystem integrals of difference densities are shown in Table~\ref{table1} and demonstrate that for the system at $r = 9$\,\AA{} , all excitations in the energy-range of the coupled $Q_x$ excitations and the higher energy dark states exhibit charge-transfer character. We classify $E_4$, which is in the energy range of the coupled $Q_x$ excitations and corresponds to transfer of half an electron from P$_{A1}$ to P$_{A2}$ as a partial charge-transfer state (PCT) in Figure~\ref{fig6}a. Our results are qualitatively similar when using the $GW$+BSE approach, as shown in Figure~S6 and consistent with our discussion in Section~\ref{sec:real-dimer}. 

These trends are even more apparent in Figure~\ref{fig6}b, where we plot the energy of all dark excitations as a function of distance and indicate their charge-transfer character in color. In the energy range considered here, there are four dark excitations without charge-transfer character which are essentially independent of distance and are only redshifted and acquire substantial charge-transfer character at relatively small $r$. The two charge-transfer states exhibit a significant distance dependence and are red-shifted by almost 1\,eV with decreasing $r$ but lose some of their charge-transfer character at the smallest distance where they start mixing with the coupled $Q_x$ excitations.
\begin{table}[htb]
\centering
\begin{tabular}[t]{c|c|cccccc}
\toprule
   $r$ (\AA)  &  molecule & \multicolumn{5}{c}{charge distribution}\\
 &&E$_4$& E$_5$& E$_6$& E$_7$&E$_8$ \\
\midrule                
   \multirow{2}{*}{9 }& P$_{A1}$ &-0.48 &0.21 &0.28 &-0.62&-0.63\\
                         & P$_{A2}$ &0.48 &-0.21 &-0.28 &0.62&0.63 \\\midrule
   \multirow{2}{*}{11 }& P$_{A1}$ &0 &-0.96 &0.96 &0&0 \\
                         & P$_{A2}$ &0 &0.96 &-0.96&0  &0\\\midrule
    \multirow{2}{*}{13 } & P$_{A1}$ & 0& 0 &0 &-0.99&0.99 \\
                           & P$_{A2}$ & 0&0 &0 &0.99&-0.99 \\\midrule                     
\bottomrule
	\end{tabular}
		\caption{Charge distribution on each molecule in the artificial dimer upon excitation as calculated by integration over subsystem difference densities. The first two excitations, i.e., E$_1$ and E$_2$, are not included since their subsystem difference densities integrate to zero.}
\label{table1}
\end{table}

For investigating the effect of the relative orientation of the two molecules, we fixed the intermolecular distance at 13\,\AA{}. Shorter distances were not possible due to overlap of functional groups for some orientations. Since rotations around the angles $\alpha$, $\beta$, and $\gamma$ do not commute, we treat them separately from each other, i.e., we first consider rotations around $\alpha$ for fixed $\beta$ and $\gamma$, then rotations around $\beta$ for fixed $\alpha$ and $\gamma$, and finally rotations around $\gamma$ for fixed $\alpha$ and $\beta$. For each structure, we determine the smallest intermolecular distance between every two individual atoms in $P_{A1}$ and $P_{A2}$, $R$. The difference between $R$ in the reference (unrotated) structure from each rotated structure, $\Delta R = R_{ref} - R_{rot}$, as a function of rotation angle, is shown in Figure~\ref{fig6}c. Since charge-transfer excitations CT$_1$ and CT$_2$ follow similar trends, we only show the change in energy of CT$_1$ upon rotation in Figure~\ref{fig6}c. Negative (positive) values of $\Delta E^{CT_1} = E^{CT_1}_{ref} - E^{CT_1}_{rot}$ correspond to a redshift (blue-shift) of the excitation energy.

Rotations around $\alpha$ and $\beta$ correspond to orientations with smaller $R$ than in the reference structure. Consequently, we observe increased intermolecular coupling and hence a redshift of the charge-transfer state by up to $\sim$0.2\,eV. For the structure for which we observe the largest effect (corresponding to a $\beta$ rotation of 120 degrees), it is primarily the relative orientation and distance of carbon chains determining the intermolecular coupling (Figure~S8a). For many of the other structures that show pronounced redshifts, we find that the functional groups of the two BCLs highlighted in Figure~\ref{fig5} are in close spatial proximity (see Figure~S8b for an example). In contrast, the rotation around the angle $\gamma$ results primarily in structures with positive $\Delta$R and a blueshift of the charge-transfer excitation by up to $\sim$0.1\,eV. We note that in the majority of structures rotated around $\gamma$, the functional groups FG$_1$, FG$_2$, and FG$_4$ are far apart from the second BCL. However, for some structures, overlap between FG$_2$ and the second BCL molecule led to unrealistic structures that were excluded from Figure~\ref{fig6}c. Overall, the $\gamma$ rotation primarily leads to geometries with weaker intermolecular coupling and an overall blueshift in energy of the charge-transfer excitation.

\subsection{Vibrational Effects on Charge-Transfer Excitations}
\label{sec:vib}
\begin{figure}[htb]
\centering
\includegraphics[width=\textwidth]{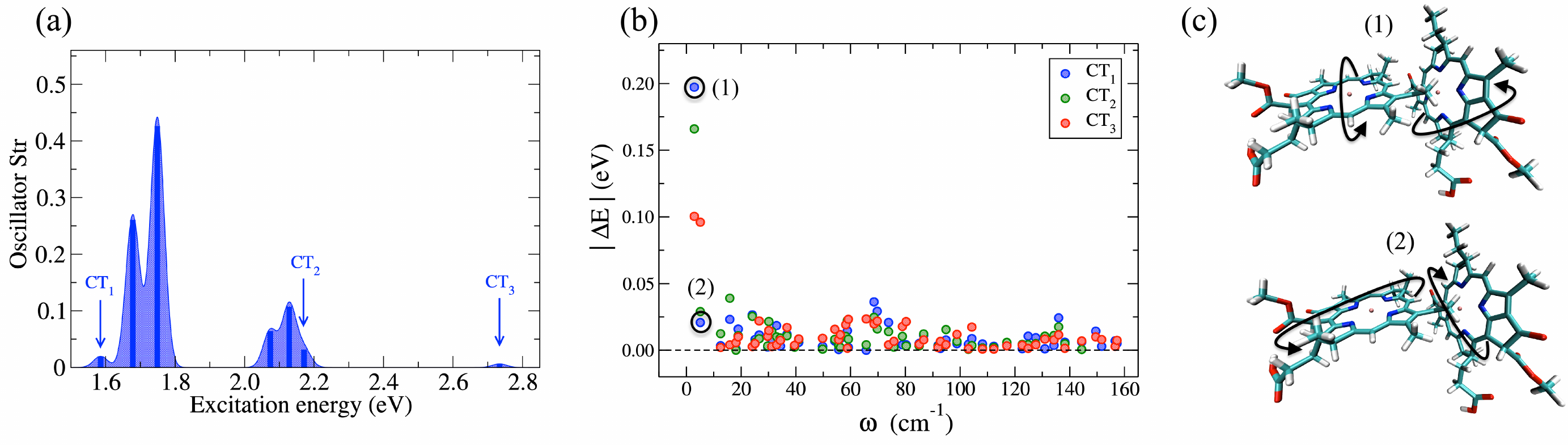}
\caption{(a) Absorption spectrum of relaxed dimer. Arrows mark the first three charge-transfer excitations, (b) Excitation energy change $|\Delta E|$ as a function of normal mode frequency for CT$_1$, CT$_2$, and CT$_3$. (c) Visualization of the first two normal modes which correspond to intermolecular rotations (see main text).}
\label{fig7}
\end{figure}

Excitations of different spatial localization and character are known to be affected in different ways by molecular vibrations \cite{alvertis_impact_2020}. Our goal here is to provide a mode-resolved picture of excitation energy changes in a BCL dimer due to thermally-activated vibrations, following earlier work by Hele \textit{et al.} \cite{hele_systematic_2021}. For this purpose, we started from the crystal structure of the special pair dimer SP and performed a full geometry optimization using the def2-TZVP basis set and B3LYP exchange-correlation functional. In the absence of the protein environment and other co-factors, no external force fixes P$_A$ and P$_B$ in the parallel configuration they have \textit{in vivo}. Consequently, the relaxed structure differs considerably from SP, and is more akin to the A-branch dimer. Since our aim is to provide a qualitative picture, we proceed with this structure which is dynamically stable, i.e., without imaginary frequencies. We note, however, that the excitation spectrum of the relaxed dimer, shown in Figure~\ref{fig7}a, differs from the spectra discussed so far. In particular, the spectrum displays a charge-transfer state CT$_1$ at $\sim$1.6\,eV (see also Table S9). This state mixes with the coupled $Q_y$ excitations and corresponds to the transfer of 0.78 of an electron from $P_A$ to $P_B$ (see Table S10). A second charge-transfer state CT$_2$ mixes with the coupled $Q_x$ excitations, while the third one, CT$_3$, is energetically well-separated from the $Q$-band excitations at $\sim$2.7\,eV.

We calculate the vibrational normal modes of the relaxed dimer using the same basis set and exchange-correlation functional but with a very fine grid for the quadrature of the exchange-correlation energy. We then generate 60 structures by distorting the relaxed reference structure along each of the 60 lowest-frequency normal modes. We assume a classical distribution to approximate the amplitude of these distortions at T = 300\,K and generate one distorted structure per mode corresponding to a positive distortion amplitude. The excitation spectrum of each distorted structure is then calculated with TDDFT as before, i.e., with $\omega$PBE with $\omega=0.171$\,$a_0^{-1}$. We define the excitation energy change of excitation $n$ as $\Delta E^{n} = E^{n}_{ref} - E^{n}_{dis}$. Here we focus on how molecular vibrations affect charge-transfer excitations, but note that $\Delta E$ for the coupled $Q_y$ and $Q_x$ excitations can also be substantial as shown in Figure~S9. 

While $\Delta E$ can be either positive or negative, depending on the direction of the distortion mode, we show $|\Delta E|$ of the charge-transfer excitations CT$_1$, CT$_2$, and CT$_3$ in Figure~\ref{fig7}b. We find that high-frequency modes correspond to intramolecular vibrations such as C-C and C-H stretch modes, which are not thermally activated and only have a small effect on the energy of the three charge-transfer states. In contrast, low-frequency modes correspond to intermolecular vibrations that change the orbital overlap between neighboring molecules and thus have a more substantial impact. In particular, we find that the two lowest-frequency modes lead to substantial changes in all three charge-transfer states. Both modes correspond to a rotational motion of the porphyrin planes of the BCL molecules with respect to each other as indicated in Figure~\ref{fig7}c. The first mode leads to a redshift of all three excitations which is with $\sim$0.2\,eV most pronounced for CT$_1$, the second one leads to a smaller blueshift of CT$_1$ and CT$_3$ and a slight redshift of CT$_2$. These results qualitatively agree with our results in Section~\ref{sec:art-dimer}, suggesting that thermally-activated vibrational modes can significantly affect the energy of charge-transfer excitations affecting their charge-transfer character and mixing with other delocalized and localized excitations of the system.

\section{Summary and Conclusions}
In summary, we have presented a systematic first-principles study of charge-transfer excitations in BCL dimers. Our model systems are inspired by molecular aggregates found in the LHII complex and RC of purple bacteria and cover a wide range of intermolecular coupling strengths, and consequently, excited-state structures. Charge-transfer excitations can be found in a wide range of energies, primarily depending on intermolecular distance and orientation. BCL molecules have a complex three-dimensional structure with several functional groups, a long phytyl tail, and other carbon chains protruding out of the porphyrin plane. \textit{In vivo}, i.e., within the evolutionary-optimized protein networks of the photosynthetic apparatus, the protein environment determines the distance, orientation, and structural details of these aggregates. Furthermore, the protein environment indirectly affects the excited state structure and dynamics of BCL aggregates through dielectric screening and electrostatic effects \cite{Stanley1996,Steffen1994,Hiyama2011,Lockhart1990,Steffen1994,Alden1995,Gunner1996,Saggu2019,Tamura2021,Brutting2021, volpert_delocalized_2022}. Therefore our results can not directly be used to infer charge-transfer mechanisms in photosynthetic systems Nonetheless, they provide an intuitive understanding and design rules for tailoring charge-transfer excitations in BCLs and similar photoactive molecules. Furthermore, they explicitly confirm the importance of charge-transfer excitations for a correct description of the $Q$-band excitations of BCL aggregates \cite{Li2017a}. We hope that our results inspire future calculations of the excited-state structure and dynamics of pigment-protein complexes and chromophore aggregates based on model Hamiltonians, that include charge-transfer excitations. 

Furthermore, we have compared our results based on TDDFT with the optimally-tuned $\omega$PBE functional to calculations using the $GW$+BSE approach. While charge-transfer excitations appear at very similar energies with both approaches, coupled $Q_y$ excitations are systematically overestimated by TDDFT as compared to the $GW$+BSE approach. Previous studies suggest that $Q_y$ excitation energies from $GW$+BSE are in better agreement with wavefunction-based methods and experiment than TDDFT with $\omega$PBE \cite{Hashemi2021a, forster_quasiparticle_2022}. However, accurate benchmarks for larger molecular aggregates are missing and we therefore do not think that a clear recommendation for using $GW$+BSE instead of TDDFT is warranted. Nonetheless, with advances in code implementation \cite{Bruneval2016,Foerster2020,duchemin_cubic-scaling_2021} and in the combination of $GW$+BSE with discrete and polarizable continuum models \cite{Duchemin2016b, Duchemin2018} and other QM/MM methods \cite{Wehner2018}, $GW$+BSE calculations of large molecular aggregates are becoming computationally feasible, demonstrated in a recent study by Förster \textit{et al.} \cite{forster_quasiparticle_2022}. Further study of the accuracy and predictive power of TDDFT, with exchange-correlation functionals that capture the nonlocal nature of charge-transfer excitations for such aggregates is necessary.

\section*{Supplementary Material}
Additional convergence data, excitation energies, difference densities and transition densities not shown in the main text, and structure files.

\section*{Acknowledgements}
This work was supported by the Bavarian State Ministry of Science and the Arts through the Elite Network Bavaria (ENB) and through computational resources provided by the Bavarian Polymer Institute (BPI).

\section*{References}\providecommand{\newblock}{}


\begin{thebibliography}{101}
\expandafter\ifx\csname url\endcsname\relax
  \def\url#1{{\tt #1}}\fi
\expandafter\ifx\csname urlprefix\endcsname\relax\def\urlprefix{URL }\fi
\providecommand{\eprint}[2][]{\url{#2}}

\bibitem{wahadoszamen_role_2014}
Wahadoszamen M, Margalit I, Ara A~M, van Grondelle R and Noy D 2014 {\em Nature
  Comm.\/} {\bf 5} 5287\urlprefix\url{https://www.nature.com/articles/ncomms6287}

\bibitem{zoppi_charge-transfer_2018}
Zoppi L and Baldridge K~K 2018 {\em Int. J. Quant.
  Chem.\/} {\bf 118} e25413
  \urlprefix\url{https://onlinelibrary.wiley.com/doi/abs/10.1002/qua.25413}

\bibitem{muntwiler_coulomb_2008}
Muntwiler M, Yang Q, Tisdale W~A and Zhu X~Y 2008 {\em Phys. Rev.
  Lett.\/} {\bf 101} 196403
  \urlprefix\url{https://link.aps.org/doi/10.1103/PhysRevLett.101.196403}

\bibitem{ohkita_charge_2008}
Ohkita H, Cook S, Astuti Y, Duffy W, Tierney S, Zhang W, Heeney M, McCulloch I,
  Nelson J, Bradley D~D~C and Durrant J~R 2008 {\em J. Am.
  Chem. Soc.\/} {\bf 130} 3030--3042 \urlprefix\url{https://doi.org/10.1021/ja076568q}

\bibitem{pensack_barrierless_2009}
Pensack R~D and Asbury J~B 2009 {\em J. Am. Chem.
  Soc.\/} {\bf 131} 15986--15987 \urlprefix\url{https://doi.org/10.1021/ja906293q}

\bibitem{lee_charge_2010}
Lee J, Vandewal K, Yost S~R, Bahlke M~E, Goris L, Baldo M~A, Manca J~V and
  Van~Voorhis T 2010 {\em J. Am. Chem. Soc.\/} {\bf 132}
  11878--11880
  \urlprefix\url{https://doi.org/10.1021/ja1045742}

\bibitem{bakulin_role_2012}
Bakulin A~A, Rao A, Pavelyev V~G, van Loosdrecht P~H~M, Pshenichnikov M~S,
  Niedzialek D, Cornil J, Beljonne D and Friend R~H 2012 {\em Science\/} {\bf
  335} 1340--1344\urlprefix\url{https://www.science.org/doi/10.1126/science.1217745}

\bibitem{caruso_long-range_2012}
Caruso D and Troisi A 2012 {\em Proc. Natl. Acad.
  Sci.\/} {\bf 109} 13498--13502
  \urlprefix\url{https://www.pnas.org/doi/full/10.1073/pnas.1206172109}

\bibitem{murthy_mechanism_2012}
Murthy D~H~K, Gao M, Vermeulen M~J~W, Siebbeles L~D~A and Savenije T~J 2012
  {\em The J. of Phys. Chem. C\/} {\bf 116} 9214--9220
  \urlprefix\url{https://doi.org/10.1021/jp3007014}

\bibitem{yost_electrostatic_2013}
Yost S~R and Van~Voorhis T 2013 {\em The J. of Phys. Chem. C\/}
  {\bf 117} 5617--5625
  \urlprefix\url{https://doi.org/10.1021/jp3125186}

\bibitem{jakowetz_what_2016}
Jakowetz A~C, Böhm M~L, Zhang J, Sadhanala A, Huettner S, Bakulin A~A, Rao A
  and Friend R~H 2016 {\em J. Am. Chem. Soc.\/} {\bf
  138} 11672--11679
  \urlprefix\url{https://doi.org/10.1021/jacs.6b05131}

\bibitem{lee_higher-energy_2017}
Lee D, Forsuelo M~A, Kocherzhenko A~A and Whaley K~B 2017 {\em The J. of
  Phys. Chem. C\/} {\bf 121} 13043--13051
  \urlprefix\url{https://doi.org/10.1021/acs.jpcc.7b03197}

\bibitem{Jordanides2001}
Jordanides X~J, Scholes G~D and Fleming G~R 2001 {\em J. Phys. Chem. B\/} {\bf
  105} 1652--1669

\bibitem{camara-artigas_interactions_2002}
Camara-Artigas A, Brune D and Allen J~P 2002 {\em Proc. Natl.
  Acad. Sci.\/} {\bf 99} 11055--11060
  \urlprefix\url{https://doi.org/10.1073/pnas.162368399}

\bibitem{Jonas1996}
Jonas D~M, Lang M~J, Nagasawa Y, Joo T and Fleming G~R 1996 {\em J. Phys.
  Chem.\/} {\bf 100} 12660--12673

\bibitem{Schlau-Cohen2012}
Schlau-Cohen G~S, Re E~D, Cogdell R~J and Fleming G~R 2012 {\em J. Phys. Chem.
  Lett.\/} {\bf 3} 2487--2492

\bibitem{Rancova2016}
Rancova O, Jankowiak R, Kell A, Jassas M and Abramavicius D 2016 {\em J. Phys.
  Chem. B\/} {\bf 120} 5601--5616

\bibitem{mirkovic2017light}
Mirkovic T, Ostroumov E~E, Anna J~M, Van~Grondelle R and Scholes G~D 2017 {\em
  Chem. Rev.s\/} {\bf 117} 249--293

\bibitem{Niedringhaus2018}
Niedringhaus A, Policht V~R, Sechrist R, Konar A, Laible P~D, Bocian D~F,
  Holten D, Kirmaier C and Ogilvie J~P 2018 {\em Proc. Nat. Acad. Sci.\/} {\bf
  115} 3563--3568

\bibitem{kawashima_energetic_2018}
Kawashima K and Ishikita H 2018 {\em Chem. Science\/} {\bf 9} 4083--4092
  ISSN 2041-6520, 2041-6539
  \urlprefix\url{http://xlink.rsc.org/?DOI=C8SC00424B}

\bibitem{Kavanagh2020}
Kavanagh M~A, Karlsson J~K, Colburn J~D, Barter L~M and Gould I~R 2020 {\em
  Proc. Nat. Acad. Sci.\/} {\bf 117} 19705--19712

\bibitem{cupellini2020successes}
Cupellini L, Bondanza M, Nottoli M and Mennucci B 2020 {\em Biochimica et
  Biophysica Acta (BBA)-Bioenergetics\/} {\bf 1861} 148049

\bibitem{wraight_absolute_1974}
Wraight C~A and Clayton R~K 1974 {\em Biochimica et Biophysica Acta (BBA) -
  Bioenergetics\/} {\bf 333} 246--260
  \urlprefix\url{https://www.sciencedirect.com/science/article/pii/0005272874900097}

\bibitem{kirmaier_picosecond-photodichroism_1985}
Kirmaier C, Holten D and Parson W~W 1985 {\em Biochimica et Biophysica Acta
  (BBA) - Bioenergetics\/} {\bf 810} 49--61
  \urlprefix\url{https://www.sciencedirect.com/science/article/pii/0005272885902051}

\bibitem{zinth_first_2005}
Zinth W and Wachtveitl J 2005 {\em ChemPhysChem\/} {\bf 6} 871--880
  \urlprefix\url{https://onlinelibrary.wiley.com/doi/abs/10.1002/cphc.200400458}

\bibitem{ma_vibronic_2018}
Ma F, Romero E, Jones M~R, Novoderezhkin V~I and van Grondelle R 2018 {\em J. Phys. Chem. Lett.\/} {\bf 9}
  \urlprefix\url{https://doi.org/10.1021/acs.jpclett.8b00108}

\bibitem{Ma2019}
Ma F, Romero E, Jones M~R, Novoderezhkin V~I and van Grondelle R 2019 {\em
  Nature Comm.\/} {\bf 10} 933
    \urlprefix\url{https://www.nature.com/articles/s41467-019-08751-8}

\bibitem{policht_hidden_2022}
Policht V~R, Niedringhaus A, Willow R, Laible P~D, Bocian D~F, Kirmaier C,
  Holten D, Mančal T and Ogilvie J~P 2022 {\em Science Advances\/} {\bf 8}
  eabk0953
  \urlprefix\url{https://www.science.org/doi/10.1126/sciadv.abk0953}

\bibitem{van_brederode_efficiency_1998}
van Brederode M~E, Ridge J~P, van Stokkum I~H~M, van Mourik F, Jones M~R and
  van Grondelle R 1998 {\em Photosyth. Res.\/} {\bf 55} 141--146
  \urlprefix\url{https://doi.org/10.1023/A:1005925917867}

\bibitem{zhou_probing_1998}
Zhou H and Boxer S~G 1998 {\em J. Phys. Chem. B\/} {\bf 102}
  9139--9147
  \urlprefix\url{https://doi.org/10.1021/jp982043w}

\bibitem{lin_excitation_1998}
Lin S, Jackson J, Taguchi A~K~W and Woodbury N~W 1998 {\em J. Phys. Chem. B\/} {\bf 102} 4016--4022 \urlprefix\url{https://doi.org/10.1021/jp980360x}

\bibitem{huang_cofactor-specific_2012}
Huang L, Ponomarenko N, Wiederrecht G~P and Tiede D~M 2012 {\em Proc.
  Natl. Acad. Sci.\/} {\bf 109} 4851--4856
  \urlprefix\url{https://www.pnas.org/doi/10.1073/pnas.1116862109}

\bibitem{fassioli_photosynthetic_2014}
Fassioli F, Dinshaw R, Arpin P~C and Scholes G~D 2014 {\em J. Roy.
  Soc. Interface\/} {\bf 11} 20130901
  \urlprefix\url{https://royalSoc.publishing.org/doi/10.1098/rsif.2013.0901}

\bibitem{Jang2018}
Jang S~J and Mennucci B 2018 {\em Rev. Mod. Phys.\/} {\bf 90} 35003
  \urlprefix\url{https://doi.org/10.1103/RevModPhys.90.035003}

\bibitem{VanderVegte2015}
van~der Vegte C~P, Prajapati J~D, Kleinekathöfer U, Knoester J and Jansen
  T~L~C 2015 {\em J. Phys. Chem. B\/} {\bf 119} 1302--13
  \urlprefix\url{http://www.ncbi.nlm.nih.gov/pubmed/25554919}

\bibitem{Curutchet2016}
Curutchet C and Mennucci B 2016 {\em Chem. Rev.\/}
  \urlprefix\url{http://pubs.acs.org/doi/abs/10.1021/acs.chemrev.5b00700}

\bibitem{thyrhaug_identification_2018}
Thyrhaug E, Tempelaar R, Alcocer M~J~P, Žídek K, Bína D, Knoester J, Jansen
  T~L~C and Zigmantas D 2018 {\em Nature Chem.\/} {\bf 10} 780--786
  \urlprefix\url{https://www.nature.com/articles/s41557-018-0060-5}

\bibitem{voityuk_fragment_2014}
Voityuk A~A 2014 {\em J. Chem. Phys.\/} {\bf 140} 244117
  \urlprefix\url{https://aip.scitation.org/doi/full/10.1063/1.4884944}

\bibitem{voityuk_interaction_2015}
Voityuk A~A 2015 {\em J. Phys. Chem. B\/} {\bf 119}
  7417--7421
  \urlprefix\url{https://doi.org/10.1021/jp511035p}

\bibitem{Li2017a}
Li X, Parrish R~M, Liu F, Kokkila~Schumacher S~I and Martínez T~J 2017 {\em J.
  Chem. Theor. Comp.\/} {\bf 13} 3493--3504

\bibitem{sen_understanding_2021}
Sen S, Mascoli V, Liguori N, Croce R and Visscher L 2021 {\em J.
  Phys. Chem. A\/} {\bf 125} 4313--4322
  \urlprefix\url{https://doi.org/10.1021/acs.jpca.1c01467}

\bibitem{Dreuw2004}
Dreuw A and Head-Gordon M 2004 {\em J. Am. Chem. Soc.\/} {\bf 126} 4007--4016

\bibitem{Refaely-Abramson2011}
Refaely-Abramson S, Baer R and Kronik L 2011 {\em Phys. Rev. B\/} {\bf 84}
  075144

\bibitem{Refaely-Abramson2012}
Refaely-Abramson S, Sharifzadeh S, Govind N, Autschbach J, Neaton J~B, Baer R
  and Kronik L 2012 {\em Phys. Rev. Lett.\/} {\bf 109} 226405
  \urlprefix\url{http://link.aps.org/doi/10.1103/PhysRevLett.109.226405}

\bibitem{Korzdorfer2012}
Körzdörfer T and Marom N 2012 {\em Phys. Rev. B\/} {\bf 86} 041110
\urlprefix\url{http://link.aps.org/doi/10.1103/PhysRevB.86.041110}

\bibitem{Refaely-Abramson2013}
Refaely-Abramson S, Sharifzadeh S, Jain M, Baer R, Neaton J~B and Kronik L 2013
  {\em Phys. Rev. B\/} {\bf 88} 081204
  \urlprefix\url{http://link.aps.org/doi/10.1103/PhysRevB.88.081204}

\bibitem{DeQueiroz2014}
De~Queiroz T~B and Kümmel S 2014 {\em J. Chem. Phys.\/} {\bf 141} 084303
\urlprefix\url{http://dx.doi.org/10.1063/1.4892937}

\bibitem{manna_quantitative_2018}
Manna A~K, Refaely-Abramson S, Reilly A~M, Tkatchenko A, Neaton J~B and Kronik
  L 2018 {\em J. Chem. Theo. Comput.\/} {\bf 14} 2919--2929
  ISSN 1549-9618 publisher: American Chem. Soc.
  \urlprefix\url{https://doi.org/10.1021/acs.jctc.7b01058}

\bibitem{Wing2020a}
Wing D, Ohad G, Haber J~B, Filip M~R, Gant S~E, Neaton J~B and Kronik L 2020
  {\em Proc. Nat. Acad. Sci.\/} {\bf 118} e2104556118
  \urlprefix\url{http://arxiv.org/abs/2012.03278}

\bibitem{Seidl1996}
Seidl A, Görling A, Vogl P, Majewski J~A and Levy M 1996 {\em Phys. Rev. B\/}
  {\bf 53} 3764--3774
  \urlprefix\url{http://www.ncbi.nlm.nih.gov/pubmed/9983927}

\bibitem{Kuemmel2017}
Kümmel S 2017 {\em Adv. Energy Mater.\/} {\bf 7} 1700440

\bibitem{bhandari_fundamental_2018}
Bhandari S, Cheung M~S, Geva E, Kronik L and Dunietz B~D 2018 {\em J.
  Chem. Theo. Comput.\/} {\bf 14} 6287--6294 ISSN 1549-9618
  publisher: American Chem. Soc.
  \urlprefix\url{https://doi.org/10.1021/acs.jctc.8b00876}

\bibitem{Rohlfing1998}
Rohlfing M and Louie S~G 1998 {\em Phys. Rev. Lett.\/} {\bf 81} 2312--2315

\bibitem{Rohlfing2000}
Rohlfing M and Louie S~G 2000 {\em Phys. Rev. B\/} {\bf 62} 4927
  \urlprefix\url{http://link.aps.org/doi/10.1103/PhysRevB.62.4927}

\bibitem{VanSetten2015}
van Setten M~J, Caruso F, Sharifzadeh S, Ren X, Scheffler M, Liu F, Lischner J,
  Lin L, Deslippe J~R, Louie S~G, Yang C, Weigend F, Neaton J~B, Evers F and
  Rinke P 2015 {\em J. Chem. Theory Comput.\/} {\bf 11} 5665--5687
  \urlprefix\url{http://pubs.acs.org/doi/10.1021/acs.jctc.5b00453}

\bibitem{Bruneval2015}
Bruneval F, Hamed S~M and Neaton J~B 2015 {\em J. Chem. Phys.\/} {\bf 142}
  244101

\bibitem{Rangel2017a}
Rangel T, Hamed S~M, Bruneval F and Neaton J~B 2017 {\em J. Chem. Phys.\/} {\bf
  146} 194108 \urlprefix\url{http://dx.doi.org/10.1063/1.4983126}

\bibitem{Duchemin2012}
Duchemin I, Deutsch T and Blase X 2012 {\em Phys. Rev. Lett.\/} {\bf 109}
  167801

\bibitem{duchemin_resonant_2013}
Duchemin I and Blase X 2013 {\em Phys. Rev. B\/} {\bf 87} 245412
  \urlprefix\url{https://link.aps.org/doi/10.1103/PhysRevB.87.245412}

\bibitem{Sharifzadeh2013}
Sharifzadeh S, Darancet P, Kronik L and Neaton J~B 2013 {\em J. Phys. Chem.
  Lett.\/} {\bf 4} 2197
  \urlprefix\url{http://pubs.acs.org/doi/abs/10.1021/jz401069f}

\bibitem{Blase2018}
Blase X, Duchemin I and Jacquemin D 2018 {\em Chem. Soc. Rev.\/} {\bf 47}
  1022--1043

\bibitem{forster_quasiparticle_2022}
Förster A and Visscher L 2022 {\em J. Chem. Theo.
  Comput.\/} {\bf 18} 6779--6793 \urlprefix\url{https://doi.org/10.1021/acs.jctc.2c00531}

\bibitem{Hashemi2021a}
Hashemi Z and Leppert L 2021 {\em J. Phys. Chem. A\/} {\bf 125} 2163--2172

\bibitem{volpert_delocalized_2022}
Volpert S, Hashemi Z, Foerster J~M, Marques M~R~G, Schelter I, Kümmel S and
  Leppert L 2023 {\em J. Chem. Phys.\/} just accepted manuscript \urlprefix\url{https://aip.scitation.org/doi/10.1063/5.0139691}

\bibitem{aksu2019explaining}
Aksu H, Schubert A, Geva E and Dunietz B~D 2019 {\em The J. Phys.
  Chem. B\/} {\bf 123} 8970--8975

\bibitem{sirohiwal2020protein}
Sirohiwal A, Neese F and Pantazis D~A 2020 {\em J. Am.
  Chem. Soc.\/} {\bf 142} 18174--18190

\bibitem{ahlrichs1989electronic}
Ahlrichs R, B{\"a}r M, H{\"a}ser M, Horn H and K{\"o}lmel C 1989 {\em Chem.
  Phys. Lett.\/} {\bf 162} 165--169

\bibitem{bruneval2016molgw}
Bruneval F, Rangel T, Hamed S~M, Shao M, Yang C and Neaton J~B 2016 {\em
  Comp. Phys. Comm.\/} {\bf 208} 149--161

\bibitem{Onida2002}
Onida G, Reining L and Rubio A 2002 {\em Rev. Mod. Phys.\/} {\bf 74} 601

\bibitem{blase2018bethe}
Blase X, Duchemin I and Jacquemin D 2018 {\em Chem. Soc. Rev.s\/} {\bf
  47} 1022--1043

\bibitem{blase2020bethe}
Blase X, Duchemin I, Jacquemin D and Loos P~F 2020 {\em The J. of Phys.
  Chem. Lett.\/} {\bf 11} 7371--7382

\bibitem{marques2004time}
Marques M~A and Gross E~K 2004 {\em Ann. Rev. Phys. Chem.\/} {\bf
  55} 427--455

\bibitem{schelter2019assessing}
Schelter I, Foerster J~M, Gardiner A~T, Roszak A~W, Cogdell R~J, Ullmann G~M,
  de~Queiroz T~B and K{\"u}mmel S 2019 {\em J. Chem. Phys.\/}
  {\bf 151} 134114

\bibitem{Stein2010}
Stein T, Eisenberg H, Kronik L and Baer R 2010 {\em Phys. Rev. Lett.\/} {\bf
  105} 266802
  \urlprefix\url{http://link.aps.org/doi/10.1103/PhysRevLett.105.266802}

\bibitem{Korzdorfer2011a}
Körzdörfer T, Sears J~S, Sutton C and Brédas J~L 2011 {\em J. Chem. Phys.\/}
  {\bf 135} 204107

\bibitem{baumeier_frenkel_2012}
Baumeier B, Andrienko D and Rohlfing M 2012 {\em J. Chem. Theo.
  Comp.\/} {\bf 8} 2790--2795\urlprefix\url{https://doi.org/10.1021/ct300311x}

\bibitem{Godbout1992}
Godbout N, Salahub D~R, Andzelm J and Wimmer E 1992 {\em Can. J. Chem.\/} {\bf
  70} 560--571

\bibitem{Bruneval2016a}
Bruneval F 2016 {\em J. of Chem. Physics\/} {\bf 145}
  \urlprefix\url{http://dx.doi.org/10.1063/1.4972003}

\bibitem{Plasser2014}
Plasser F, Wormit M and Dreuw A 2014 {\em J. Chem. Phys.\/} {\bf 141} 024106

\bibitem{Plasser2012}
Plasser F and Lischka H {\em J. Chem. Theor. Comput.\/} {\bf 8} 2777 -- 2789
  \urlprefix\url{http://dx.doi.org/10.1021/ct300307c}

\bibitem{camara2002interactions}
Camara-Artigas A, Brune D and Allen J 2002 {\em Proc. Natl.
  Acad. Sci.\/} {\bf 99} 11055--11060

\bibitem{papiz2003structure}
Papiz M~Z, Prince S~M, Howard T, Cogdell R~J and Isaacs N~W 2003 {\em J.
  Mol. Biol.\/} {\bf 326} 1523--1538

\bibitem{Dreuw2010}
Dreuw A, Harbach P~H, Mewes J~M and Wormit M 2010 {\em Theo. Chem.
  Acc.\/} {\bf 125} 419--426
  
\bibitem{kasha1965exciton}
Kasha M, Rawls H~R and El-Bayoumi M~A 1965 {\em Pure Appl. Chem.\/}
  {\bf 11} 371--392

\bibitem{alvertis_impact_2020}
Alvertis A~M, Pandya R, Muscarella L~A, Sawhney N, Nguyen M, Ehrler B, Rao A,
  Friend R~H, Chin A~W and Monserrat B 2020 {\em Phys. Rev. B\/} {\bf 102}
  081122
  \urlprefix\url{https://link.aps.org/doi/10.1103/PhysRevB.102.081122}

\bibitem{hele_systematic_2021}
Hele T~J~H, Monserrat B and Alvertis A~M 2021 {\em J. Chem.
  Phys.\/} {\bf 154} 244109
  \urlprefix\url{https://aip.scitation.org/doi/10.1063/5.0052247}

\bibitem{Stanley1996}
Stanley R~J, King B and Boxer S~G 1996 {\em J. Phys. Chem.\/} {\bf 100}
  12052--12059

\bibitem{Steffen1994}
Steffen M~A, Lao K and Boxer S~G 1994 {\em Science\/} {\bf 264} 810--816

\bibitem{Hiyama2011}
Hiyama M and Koga N 2011 {\em Photochem. Photobiol.\/} {\bf 87} 1297--1307

\bibitem{Lockhart1990}
Lockhart D~J, Kirmaier C, Holten D and Boxer S~G 1990 {\em J. Chem. Phys.\/}
  {\bf 94} 6987--6995

\bibitem{Alden1995}
Alden R~G, Parson W~W, Chu Z~T and Warshel A 1995 {\em J. Am. Chem. Soc.\/}
  12284--12298

\bibitem{Gunner1996}
Gunner M~R, Nicholls A and Honig B 1996 {\em J. Phys. Chem.\/} {\bf 100}
  4277--4291

\bibitem{Saggu2019}
Saggu M, Fried S~D and Boxer S~G 2019 {\em J. Phys. Chem. B\/} {\bf 123}
  1527--1536

\bibitem{Tamura2021}
Tamura H, Saito K and Ishikita H 2021 {\em Chem. Sci.\/} {\bf 12} 8131--8140

\bibitem{Brutting2021}
Brütting M, Foerster J~M and Kümmel S 2021 {\em J. Phys. Chem. B\/} {\bf 125}
  3468--3475

\bibitem{Bruneval2016}
Bruneval F, Rangel T, Hamed S~M, Shao M, Yang C and Neaton J~B 2016 {\em
  Comput. Phys. Comm.\/} {\bf 208} 149--161

\bibitem{Foerster2020}
Förster A and Visscher L 2020 {\em J. Chem. Theor. Comp.\/} {\bf 16}
  7381--7399

\bibitem{duchemin_cubic-scaling_2021}
Duchemin I and Blase X 2021 {\em J. Chem. Theo. Comput.\/}
  {\bf 17} 2383--2393
  \urlprefix\url{https://doi.org/10.1021/acs.jctc.1c00101}

\bibitem{Duchemin2016b}
Duchemin I, Jacquemin D and Blase X 2016 {\em J. Chem. Phys.\/} {\bf 144}
  164106 \urlprefix\url{http://dx.doi.org/10.1063/1.4946778}

\bibitem{Duchemin2018}
Duchemin I, Guido C~A, Jacquemin D and Blase X 2018 {\em Chem. Sci.\/} {\bf 9}
  4430--4443

\bibitem{Wehner2018}
Wehner J, Brombacher L, Brown J, Junghans C, Çaylak O, Khalak Y, Madhikar P,
  Tirimbò G and Baumeier B 2018 {\em J. Chem. Theory Comput.\/} {\bf 14}
  6253--6268
\end{thebibliography}
\end{document}